%% file: SMEFTmW.tex
\documentclass[a4paper,11pt]{article}
\pdfoutput=1 % if your are submitting a pdflatex (i.e. if you have
\usepackage{jheppub} % for details on the use of the package, please
\usepackage[utf8]{inputenc}
\usepackage[T1]{fontenc} % if needed

\usepackage{placeins} %to use \FloatBarrier to stop figures floating past a certain point

\usepackage{chngpage}

\usepackage{amsmath,amsfonts}
\usepackage{slashed}
\usepackage{feynmp}
\usepackage{float}
\usepackage{graphicx}
\usepackage{array}
\usepackage{colortbl}
\usepackage{booktabs}
\usepackage{tcolorbox}
\newcolumntype{x}[1]{>{\centering\let\newline\\\arraybackslash\hspace{0pt}}p{#1}}

\usepackage{multicol}

\usepackage{fontawesome}
\usepackage{hyperref}
\hypersetup{
     colorlinks = true,
     linkcolor = blue,
     anchorcolor = blue,
     citecolor = blue,
     filecolor = blue,
     urlcolor = blue,
     bookmarks=true,
     linktocpage=true
     }

\renewcommand{\phi}{\ensuremath{\varphi}}

\newcommand{\bpm}{\begin{pmatrix}}
\newcommand{\epm}{\end{pmatrix}}

\definecolor{lightgray}{rgb}{0.83, 0.83, 0.83}
\definecolor{lightpurp}{rgb}{0.901,0.796,0.882}

\newcommand{\lgr}{\rowcolor{lightgray} }

\usepackage[framemethod=TikZ]{mdframed}
\mdfdefinestyle{boxed}{%
    linecolor=black,
    outerlinewidth=0.5pt,
    roundcorner=5pt,
    innertopmargin=5pt,
    innerbottommargin=10pt,
    innerrightmargin=10pt,
    innerleftmargin=10pt,
    backgroundcolor=gray!05!white
    }

\title{SMEFT Analysis of \boldmath $m_W$}
\author[a]{Emanuele~Bagnaschi,}
\author[b,a,c]{John~Ellis,}
\author[d]{Maeve~Madigan,}
\author[b]{Ken Mimasu,}
\author[e,f]{Veronica~Sanz}
\author[a,b,d,g]{and Tevong~You}

\affiliation[a]{Theoretical Physics Department, CERN, CH-1211 Geneva 23, Switzerland}
\affiliation[b]{Theoretical Particle Physics and Cosmology Group, Department of Physics, \\
King's~College~London, London WC2R 2LS, UK}
\affiliation[c]{National Institute of Chemical Physics \& Biophysics, R{\" a}vala 10, 10143 Tallinn, Estonia}
\affiliation[d]{DAMTP, University of Cambridge, Wilberforce Road, Cambridge CB3 0WA, UK}
\affiliation[e]{Instituto de F{\' i}sica Corpuscular (IFIC), Universidad de Valencia-CSIC, E-46980 Valencia, Spain}
\affiliation[f]{Department of Physics and Astronomy, University of Sussex, Brighton BN1 9QH, UK}
\affiliation[g]{Cavendish Laboratory, University of Cambridge, J.J. Thomson Avenue, Cambridge CB3 0HE, UK}
\emailAdd{emanuele.bagnaschi@cern.ch}
\emailAdd{john.ellis@cern.ch}
\emailAdd{mum20@cam.ac.uk}
\emailAdd{ken.mimasu@kcl.ac.uk}
\emailAdd{veronica.sanz@uv.es}
\emailAdd{tevong.you@cern.ch}

\abstract{
We use the {\tt Fitmaker} tool to incorporate the recent CDF measurement of $m_W$ in a global fit to
electroweak, Higgs, and diboson data in the Standard Model Effective Field Theory (SMEFT) including
dimension-6 operators at linear order. We find that including
any one of the SMEFT operators ${\cal O}_{HWB}$, ${\cal O}_{HD}$, ${\cal O}_{\ell \ell}$ or
${\cal O}_{H \ell}^{(3)}$ with a non-zero coefficient could provide a
better fit than the Standard Model, with the strongest pull for ${\cal O}_{HD}$ and no tension with other electroweak precision data. We then analyse which tree-level single-field extensions of the
Standard Model could generate such operator coefficients with the appropriate
sign, and discuss the masses and couplings of these fields that best fit
the CDF measurement and other data. In particular, the global fit favours either a singlet $Z^\prime$ vector boson, a scalar electroweak triplet with zero hypercharge, or a vector electroweak triplet with unit hypercharge, followed by a singlet heavy neutral lepton, all
with masses in the multi-TeV range for unit coupling.
}
\newpage
\begin{document}

\begin{flushright}
{\small KCL-PH-TH/2022-11, CERN-TH-2022-062}
\end{flushright}
%\notoc

\maketitle
\flushbottom

\section{Introduction}
\label{sec:intro}
The Standard Model Effective Field theory (SMEFT) provides a powerful
framework for analysing possible experimental deviations from Standard
Model (SM) predictions that could be due to new physics with an energy
scale $\Lambda$ above those explored directly by current experiments~\cite{Weinberg:1979sa, Buchmuller:1985jz}.  
The leading SMEFT contributions to experimental observables appear in linear order, and are due to
interferences between SM amplitudes and those generated by dimension-6
operators. A global analysis of such linear SMEFT effects in electroweak, Higgs, diboson
and top data using the {\tt Fitmaker} tool found consistency with the SM and no significant evidence for new physics from measurements
made during run 2 of the LHC and by previous experiments~\cite{Ellis:2020unq}.~\footnote{More
recently, some evidence for new physics has been found in measurements of
flavour observables (see~\cite{LHCb:2021trn} and references therein) and the anomalous magnetic moment of the muon~\cite{Muong-2:2006rrc,Muong-2:2021ojo}, but we
do not discuss those phenomena here.} The SMEFiT collaboration subsequently made a global analysis that included SMEFT effects at quadratic order~\cite{Ethier:2021bye}, assuming that the precision electroweak data are consistent with the SM. Other global fits have been performed for various combinations of electroweak, diboson and Higgs data, e.g.,~\cite{Ellis:2018gqa, daSilvaAlmeida:2018iqo, Biekoetter:2018ypq, Falkowski:2019hvp, Ethier:2021ydt, Almeida:2021asy, Iranipour:2022iak, Dawson:2022bxd}, as well as separate fits to mainly top measurements, e.g.,~\cite{Buckley:2015lku, Hartland:2019bjb, vanBeek:2019evb, Durieux:2019rbz, Bissmann:2019gfc,  Brivio:2019ius, CMS:2020pnn, Bissmann:2020mfi, Bruggisser:2021duo, Miralles:2021dyw, Alasfar:2022zyr, Liu:2022vgo}.

\begin{figure}[t]
\centering
\includegraphics[width=0.85\textwidth]{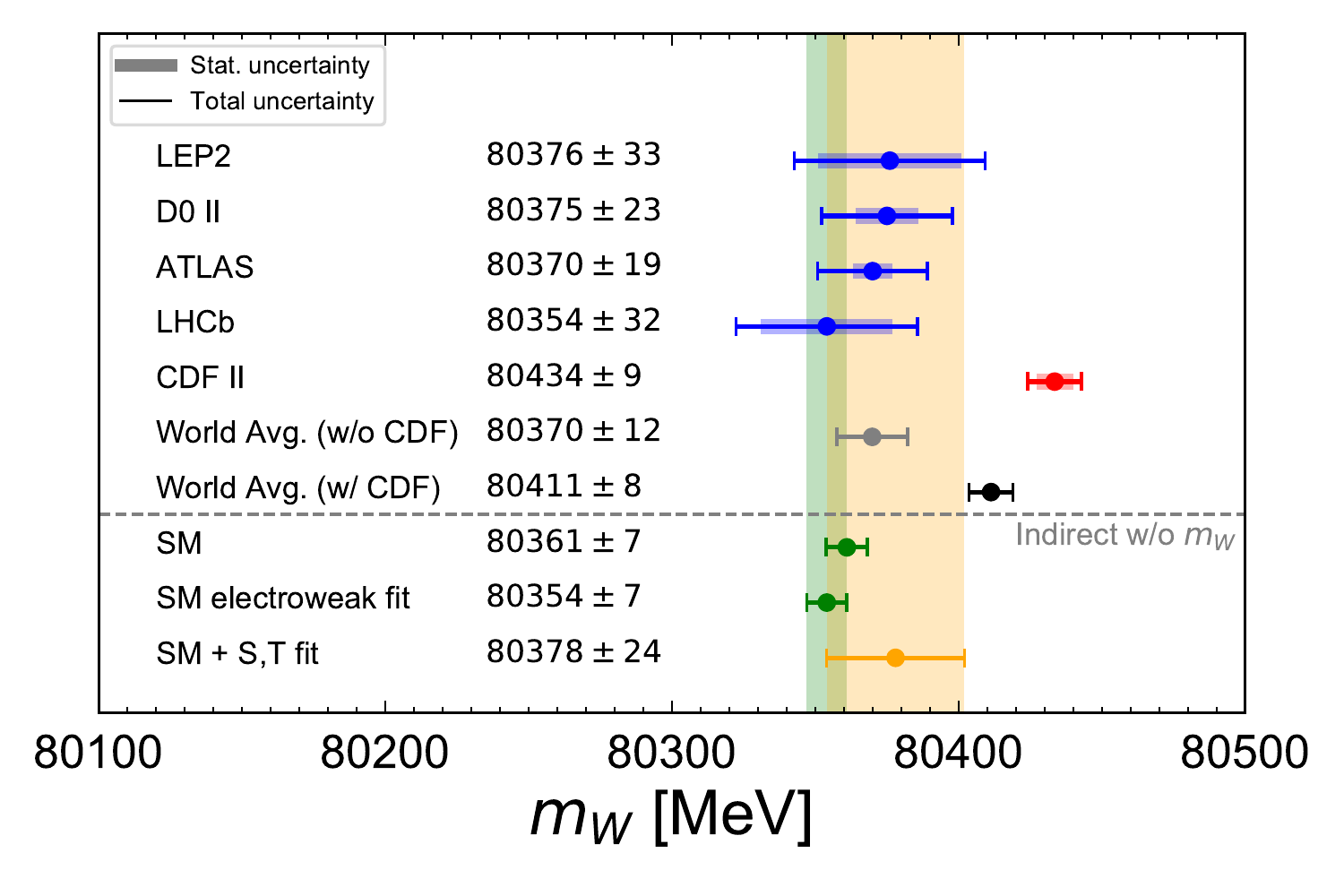}
\caption{
\label{fig:mW}
\it A comparison of the new CDF measurement of $m_W$ (red)~\cite{CDF:2022hxs} with previous measurements (blue)~\cite{ALEPH:2005ab,D0:2012kms,LHCb:2021bjt}, and a naive world average with (black) and without (grey) CDF. The values found in the Standard Model using fixed input parameters~\cite{ParticleDataGroup:2016lqr}, or determined from an electroweak fit omitting measurements of $m_W$~\cite{Haller:2018nnx}, are shown in green. The orange band is the 1-$\sigma$ range allowed in a two-parameter fit to the oblique electroweak precision parameters $S$ and $T$~\cite{Peskin:1990zt,Altarelli:1990zd} made using the SMEFT operators ${\cal O}_{HWB}$ and ${\cal O}_{HD}$ in {\tt Fitmaker}, also omitting any direct
measurement of $m_W$.}
\end{figure}

An exciting recent development has been a measurement of the $W$-boson
mass, $m_W$, by the CDF Collaboration~\cite{CDF:2022hxs}, which found
 $m_W = 80433.5 \pm 9.4$~MeV. This
value is in significant tension with the SM prediction obtained from precision
electroweak data, namely $m_W = 80354 \pm 7$~MeV~\cite{Haller:2018nnx},
and also previous direct measurements including the most precise one, that by the
ATLAS Collaboration, $m_W = 80370 \pm 19$~MeV~\cite{ATLAS:2017rzl}, as seen in Fig.~\ref{fig:mW}. Pending future scrutiny
and resolution of its tension with previous measurements, here
we accept the CDF experimental result at face value,
and use the SMEFT at linear order in dimension-6 operator coefficients to explore its
potential implications for new physics beyond the Standard Model.

There have been previous indications from LEP, Tevatron and ATLAS that $m_W$ might be slightly larger
than the SM prediction from an electroweak fit to the SM that omits the $m_W$ measurements~\cite{Haller:2018nnx},
shown as the  green band in Fig.~\ref{fig:mW}. A recent measurement by LHCb~\cite{LHCb:2021bjt} brought down the world average
of direct measurements, though its uncertainty remained relatively large. A world average obtained by
taking the combination of LEP results with D0 and CDF was given by CDF in~\cite{CDF:2022hxs}. Fig.~\ref{fig:mW}
shows in black (grey) the result of the ATLAS and LHCb
measurements combined with (omitting) the new CDF measurement neglecting correlations, displaying the apparent tension with the SM.
Some tension was evident already before the CDF measurement, and may remain even in the event that additional uncertainties are identified.
Identifying new physics able to mitigate this tension and quantifying its consistency with other data
provides a theoretical perspective that is complementary to the experimental one.

There have been many theoretical studies of the
possibility of a deviation of $m_W$ from its SM value, e.g., in the context
of extensions of the SM such as supersymmetry. However, a deviation as large as that
reported by CDF is difficult to obtain with electroweak sparticles in a minimal supersymmetric model (see~\cite{Bagnaschi:2022qhb}
and references therein). More generically, new physics parametrised by the  oblique parameters $S$ and $T$~\cite{Peskin:1990zt,Altarelli:1990zd} could accommodate a sizeable enhancement of $m_W$ while remaining compatible with electroweak precision data, as shown by the orange band in Fig.~\ref{fig:mW}. This was obtained by using  {\tt Fitmaker}
to make a fit including the SMEFT operators ${\cal O}_{HWB}$ and ${\cal O}_{HD}$ and omitting direct $m_W$ measurements.
%The SMEFT provides a general framework for analysing the types of new physics that may enter in electroweak precision physics and determinations of the $W$ mass, taking into account other measurements such as those in the Higgs, diboson and top sectors.

The main purpose of this paper is to present global fits in the general and
relatively model-independent framework provided by the SMEFT, exploring
the extent to which it can accommodate the CDF result and other measurements of $m_W$ and, if so, in what type of minimal
extension of the SM might be rersponsible. We identify several suitable single-field extensions of the
SM that can accommodate the CDF measurement and other measurements, and estimate the favoured ranges of the masses of the new particles,
finding that they
may well be sufficiently heavy for our leading-order SMEFT analysis to be
consistent. We also comment on the prospects for direct LHC searches for these new particles.
We note that the PDG has proposed~\cite{ParticleDataGroup:2016lqr}
a prescription for combining data that are only poorly consistent, and we comment on the changes
in our final results if the experimental uncertainties in Fig.~\ref{fig:mW} are rescaled using this prescription.

The CDF anomaly requires confirmation. Nevertheless, our SMEFT analysis of $M_W$ uncovers flat directions and highlights the complementarity of different datasets in constraining the multi-dimensional space of Wilson coefficients, as well as the UV extensions that they probe. Our $M_W$ analysis  represents a first step towards extending our previous SMEFT analysis to include constraints from CKM unitarity, and provides a useful guide for future phenomenological studies of models that were motivated by measurements of $M_W$ even prior to the CDF measurement~\cite{Diessner:2019ebm, Allanach:2021kzj, Alguero:2022est, Bagnaschi:2022qhb}.

\section{$m_W$ in the SMEFT}

The SMEFT Lagrangian for dimension-6 operators $\mathcal{O}_i$ has coefficients normalised as
\begin{equation}
\mathcal{L}_\text{SMEFT}^{\text{dim-6}} = \sum_{i=1}^{2499}\frac{C_i}{\Lambda^2}\mathcal{O}_i \, ,
\label{eq:Ldim6}
\end{equation}
where the $C_i$ are dimensionless Wilson coefficients and $\Lambda$ represents a dimensionful scale. The number of operators is reduced in our fit~\cite{Ellis:2020unq} by assuming a $SU(3)^5$ flavour symmetry. At linear order, four dimension-6 SMEFT operators can induce a shift in the $W$ mass, namely
\begin{eqnarray}
& {\cal O}_{HWB} \equiv H^\dagger \tau^I H\, W^I_{\mu\nu} B^{\mu\nu}, \; \; & {\cal O}_{HD} \equiv \left(H^\dagger D^\mu H\right)^\star \left(H^\dagger D_\mu H\right) \, , \nonumber \\
& {\cal O}_{\ell \ell} \equiv \left(\bar \ell_p \gamma_\mu \ell_r \right) \left(\bar \ell_s \gamma^\mu \ell_t \right), & {\cal O}_{H \ell}^{(3)} \equiv \left(H^\dagger i\,\raisebox{2mm}{\boldmath ${}^\leftrightarrow$}\hspace{-4mm} D_\mu^{\,I}\,H \right) \left(\bar \ell_p \tau^I \gamma^\mu \ell_r \right) \, ,
\label{gangof4}
\end{eqnarray}
where we adopt the Warsaw basis~\cite{Grzadkowski:2010es} for these and other dimension-6 SMEFT operators. The pole mass shift relative to the SM is given by
\begin{equation}
    \frac{\delta m_W^2}{m_W^2} = -\frac{\sin{2\theta_w}}{\cos{2\theta_w}}\frac{v^2}{4\Lambda^2}\left( \frac{\cos\theta_w}{\sin\theta_w} C_{HD} + \frac{\sin\theta_w}{\cos\theta_w}\left(4 C_{Hl}^{(3)} - 2 C_{ll}\right) + 4 C_{HWB} \right) \, .
\end{equation}
We use the electroweak input scheme that uses $\{\alpha_{EW}, G_F, M_Z\}$ as input parameters~\cite{Brivio:2021yjb}, with values
\begin{equation}
 \alpha_{EW}^{-1} = 127.95 \quad , \quad G_F = 1.16638 \times 10^{-5} \text{ GeV}^{-2} \quad , \quad m_Z = 91.1876 \text{ GeV} \, .
 \label{inputs}
\end{equation}
We neglect in our fit theoretical SMEFT errors such as a possible measurement bias in extracting the value of $m_W$ in the SMEFT, which has been shown to be negligible~\cite{Bjorn:2016zlr}.

\begin{figure}[t]
\centering
\includegraphics[width=0.5\textwidth]{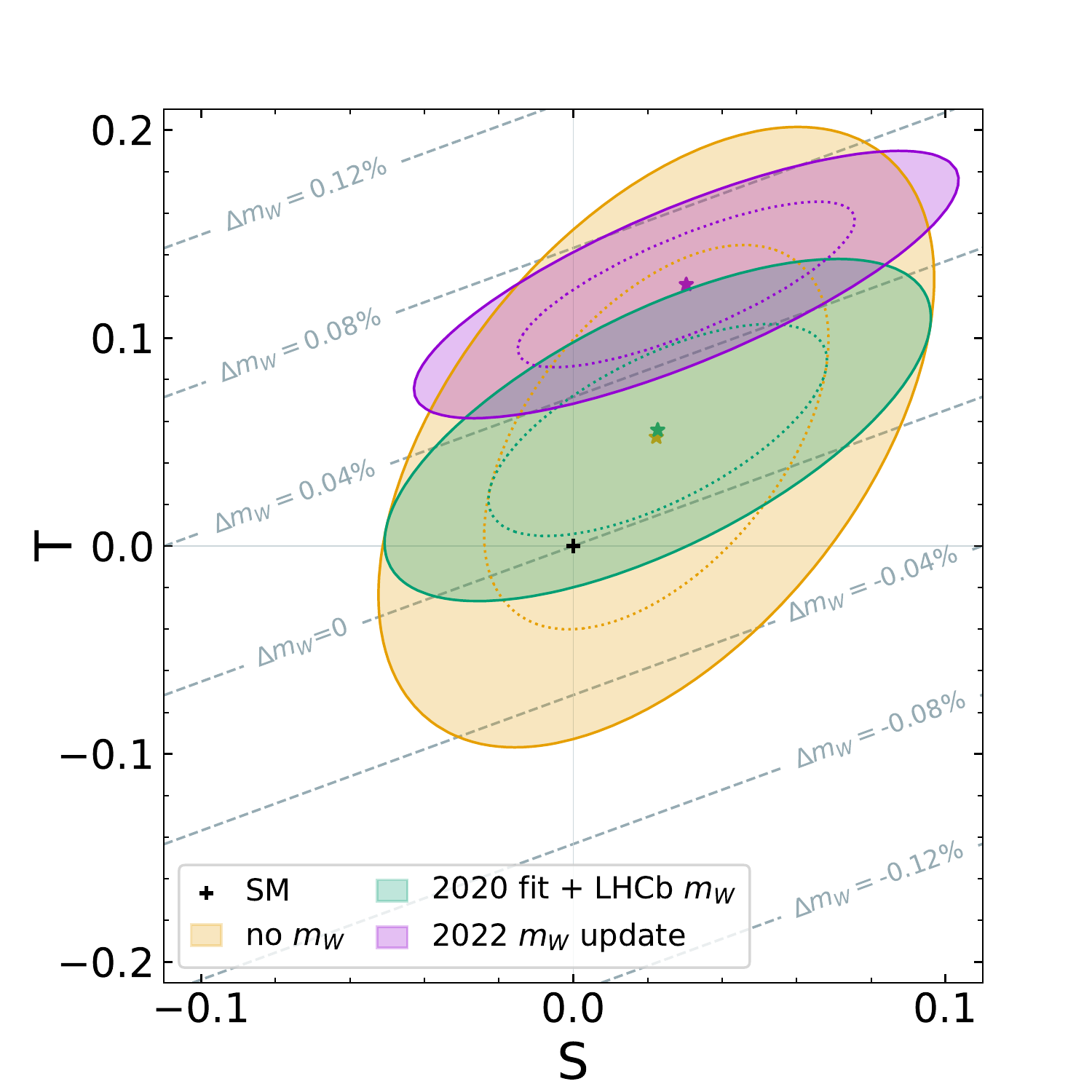}
\caption{\it The 68\% and 95\% CL contours of fits to the oblique parameters $S$ and $T$, which are equivalent to a fit including the
SMEFT operators ${\cal O}_{HWB}$ and ${\cal O}_{HD}$, see Eq.~(\ref{STSMEFT}). The orange, green, and purple contours
are the result of fits omitting $m_W$ measurements, with $m_W$ previous to the recent CDF measurement, and including the latest CDF result.
We see a clear pull away from the SM value at the origin due to the CDF result,
which nevertheless remains compatible with the range allowed by other data. The percentage shifts in the $W$ mass are denoted by dashed grey lines.
}
\label{fig:ST}
\end{figure}

A common parametrisation of new physics involves the oblique parameters $S$ and $T$~\cite{Peskin:1990zt,Altarelli:1990zd}, which can indicate the range of $m_W$ allowed by electroweak precision observables under the assumption of this 2-parameter framework. Their relation to the dimension-6 SMEFT operators in the Warsaw basis is given by
\begin{equation}
    \frac{v^2}{\Lambda^2} C_{HWB} = \frac{g_1 g_2}{16\pi} S \quad , \quad \frac{v^2}{\Lambda^2} C_{HD} = -\frac{g_1^2 g_2^2}{2\pi (g_1^2 + g_2^2)} T \, .
\label{STSMEFT}
\end{equation}
Before going into the details of our global SMEFT fit in the next Section, we show in Fig.~\ref{fig:ST} the 68\% and 95\% CL contours of fits to $S$ and $T$. The orange, green, and purple contours are the result of a fit omitting $m_W$ measurements, with $m_W$ prior to CDF's update, and including the latest CDF determination. We see a clear pull away from the SM value at the origin due to the CDF result, which nevertheless remains compatible with the range allowed by the data without $m_W$. Contours of the percent shift in $m_W$ are shown as dashed grey lines, with an enhancement of $\Delta m_W$ from 0.04\% to 0.08\% being favoured.

In the next Section we discuss the compatibility of the CDF result with the complete set of electroweak, Higgs and diboson data within
the more general SMEFT framework.

\begin{table}[t]
\renewcommand{\arraystretch}{1.1}
\begin{adjustwidth}{-1in}{-1in}% adjust the L and R margins by 1 inch
%\centering
\begin{center}
{\small
\input{limit_table}

}
\end{center}
\end{adjustwidth}
\caption{\it \label{tab:all_fit} Table of the numerical results from the global fits
to the electroweak, diboson and Higgs data in the CDF-friendly flavour-symmetric SU(3)$^5$ scenario
that are visualised in Fig.~\ref{fig:flavour_universal_EWdiBH}, switching on only each individual operator (left columns)
and including all operators and marginalising over the other operator coefficients (right columns).}
\end{table}

\begin{figure}[t]
\centering
% \includegraphics[width=1.0\textwidth]{flavour_universal_individual_fit.png}\\
%\caption{...}
%\vspace{-7mm}
\includegraphics[width=1.0\textwidth]{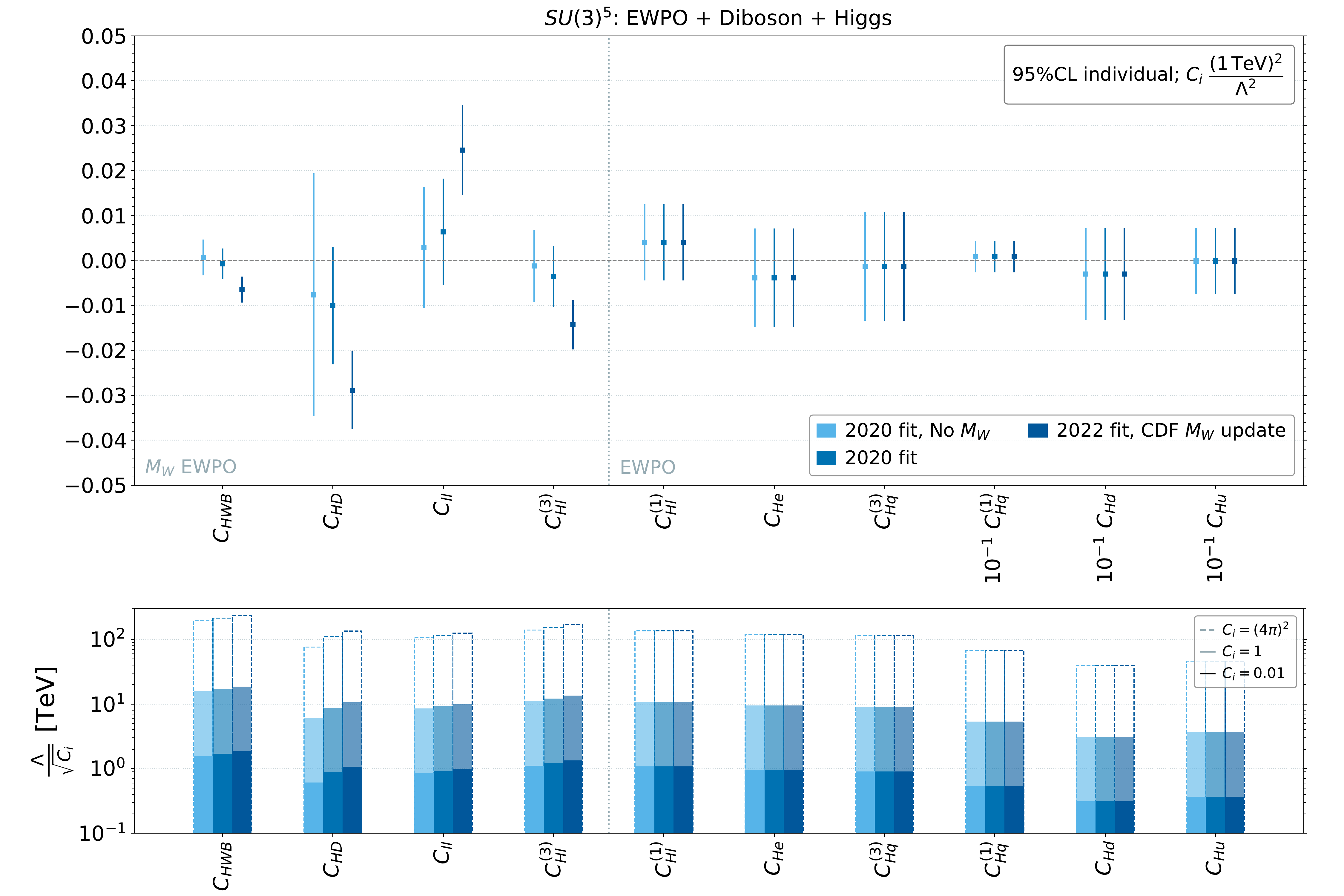}\\
%\label{fig:flavour_universal_indiv}
%\end{figure}
%
%\begin{figure}[h]
%\centering
%\vspace{-3mm}
\includegraphics[width=1.0\textwidth]{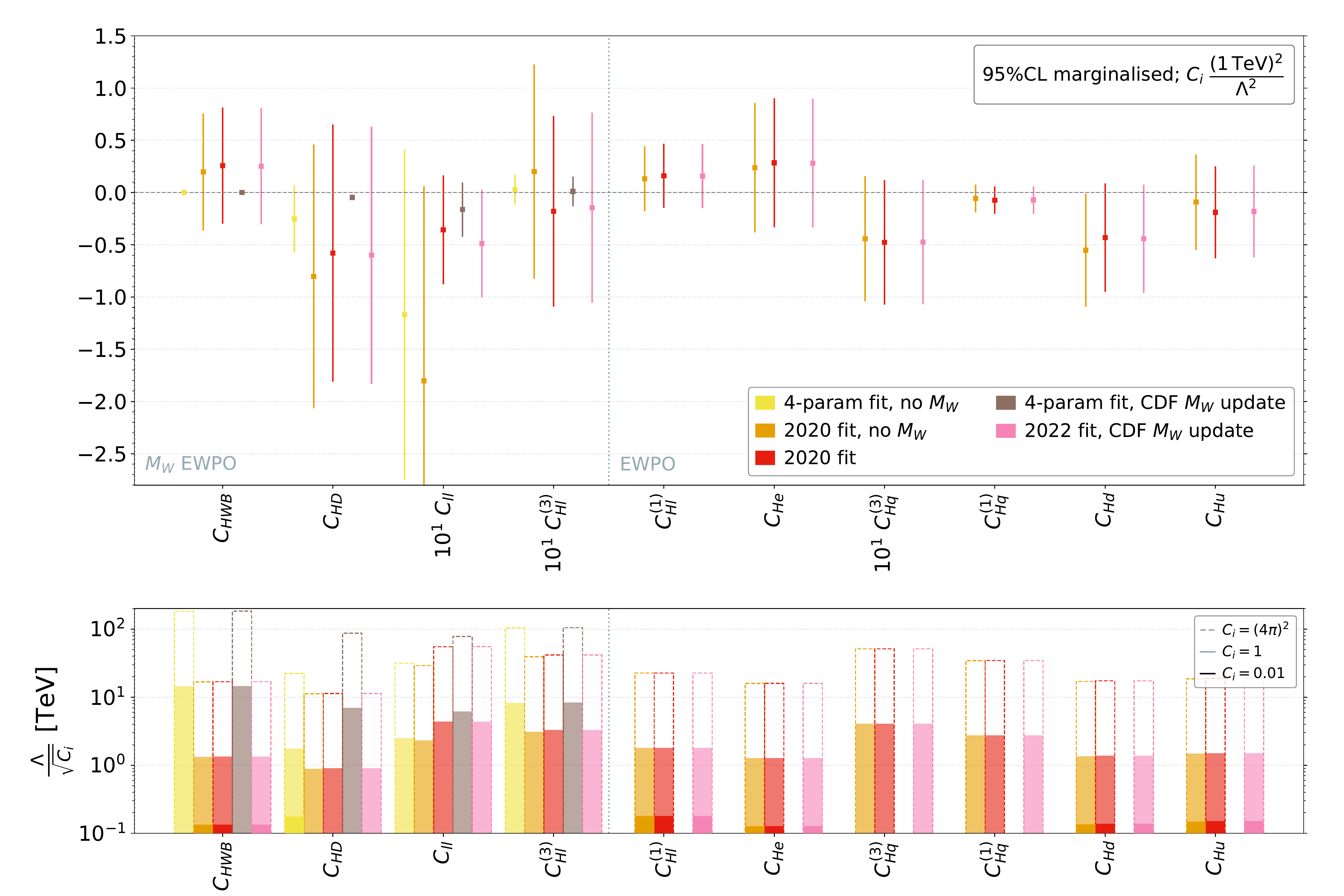}\\
\vspace{-3mm}
\caption{\it Constraints from global linear fits to measurements of $m_W$
combined with Higgs, diboson and electroweak precision observables made using {\tt Fitmaker}~\cite{Ellis:2020unq}
on the individual and marginalised operator coefficients
$C_i (1~{\rm TeV})^2/\Lambda^2$ (top and third panels, respectively)
and the corresponding scales $\Lambda$ for the indicated values of the dimensionless coupling ${C_i}$ at the 95\% confidence level
(second and bottom panels).
In the upper panels we compare results for the individual operators dropping the $m_W$ constraint,
using the $m_W$ constraint applied in~\cite{Ellis:2020unq},
and using the new CDF measurement of $m_W$.
In the lower panels, we compare the constraints obtained marginalising either over the full set of 20 operators, or over the four operators that
can modify $m_W$.
}
\label{fig:flavour_universal_EWdiBH}
\end{figure}

\clearpage
\section{SMEFT Fit Results}
\label{sec:SMEFT}

The {\tt Fitmaker} tool~\cite{Ellis:2020unq} includes
consistently all the linear (interference) effects of dimension-6 SMEFT operators.
We use the same electroweak, Higgs and diboson data set as that analysed in~\cite{Ellis:2020unq}, comparing with results
obtained by incorporating the new CDF measurement of $m_W$~\cite{CDF:2022hxs}.
We assume that the coefficients of the dimension-6 operators involving fermions are
flavour-universal, i.e., imposing on them an SU(3)$^5$ flavour symmetry and allowing
a total of 20 operators in our analysis~\footnote{Since the correlations between
top sector operators and bosonic operators were shown in~\cite{Ellis:2020unq} to be relatively weak, we expect
that including top data or making a fit in which operators involving top quarks break the flavour symmetry down to
SU(2)$^2 \times$SU(3)$^2$ would yield similar results.}.

One would expect the preferred values of the coefficients of the four
operators (\ref{gangof4}) of dimension 6 that can modify the SM prediction for $m_W$ at the
linear level, namely ${C}_{HWB, HD, \ell \ell}$ and ${C}^{(3)}_{H \ell}$, to be influenced by the inclusion of the CDF
measurement of $m_W$ in the global dataset. Indeed, we find this in global fits to
individual operator coefficients: the upper panels of Fig.~\ref{fig:flavour_universal_EWdiBH} show results for
a subset of operators most constrained by electroweak precision observables.
We see in the top panel that negative values of ${C}_{HWB}, C_{HD}$ and ${C}^{(3)}_{H \ell}$ are preferred,
but a positive value of $C_{\ell \ell}$. Comparing with the fit omitting the measurements of $m_W$ and the fit
including the old determination of $m_W$, the new best fit values remain compatible at the 95\% CL
while exhibiting a clear pull away from zero. The bars in the second panel show the sensitivities
(i.e. the widths of the 95\% CL uncertainties centred around zero)
of the respective fits to the mass scales in the coefficients of the respective operators.
The dark, light, and transparent shadings correspond to $C_i = 0.01, 1, (4\pi)^2$,  respectively.

\FloatBarrier

\begin{figure}[t!]
\centering
\includegraphics[width=0.85\textwidth]{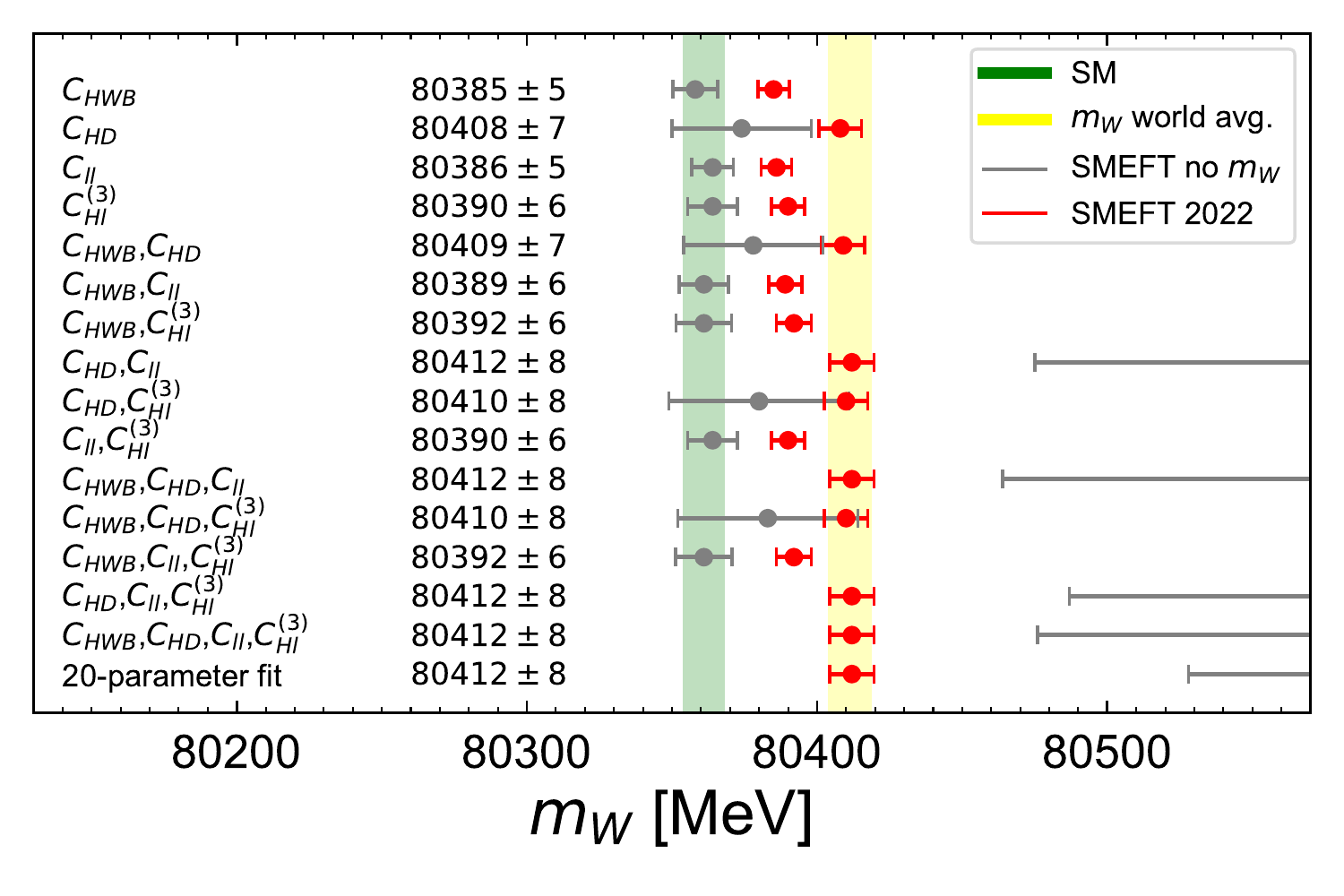}
\caption{
\label{fig:SMEFTmW}
\it Values of $m_W$ in fits including all combinations of operators entering linearly in $m_W$, including from 1 to 4 operator coefficients, as well as a fit to 20 operator coefficients. Results omitting direct measurements of $m_W$ are shown in grey, and results using the
current world average of $m_W$ measurements including that by CDF are shown in red. The vertical green band is
the SM prediction for $m_W$ based on other data, and the yellow band is the current world average of $m_W$ measurements.}
\end{figure}

On the other hand, the effects of including the
CDF measurement are less apparent in the lower panels of
Fig.~\ref{fig:flavour_universal_EWdiBH}, where we display the weaker constraints on the
operator coefficients obtained when either all operators are included in the fit and their
coefficients are marginalised over, or just the four operators that contribute to $m_W$ are included and marginalised over.
Although results for only a subset of operators most strongly constrained in electroweak precision observables are displayed in Fig.~\ref{fig:flavour_universal_EWdiBH}, all the 20 operators are included in our flavour-universal SU(3)$^5$ fit.
The effect of $m_W$ on the operators not displayed is negligible. We present in Table~\ref{tab:all_fit}
%in the Appendix
our numerical results for the 20 SU(3)$^5$-symmetric dimension-6 operator coefficients
in the individual and fully marginalised fits when
the CDF value of $m_W$ is included.

\begin{figure}[t!]
\centering
\includegraphics[width=0.9\textwidth]{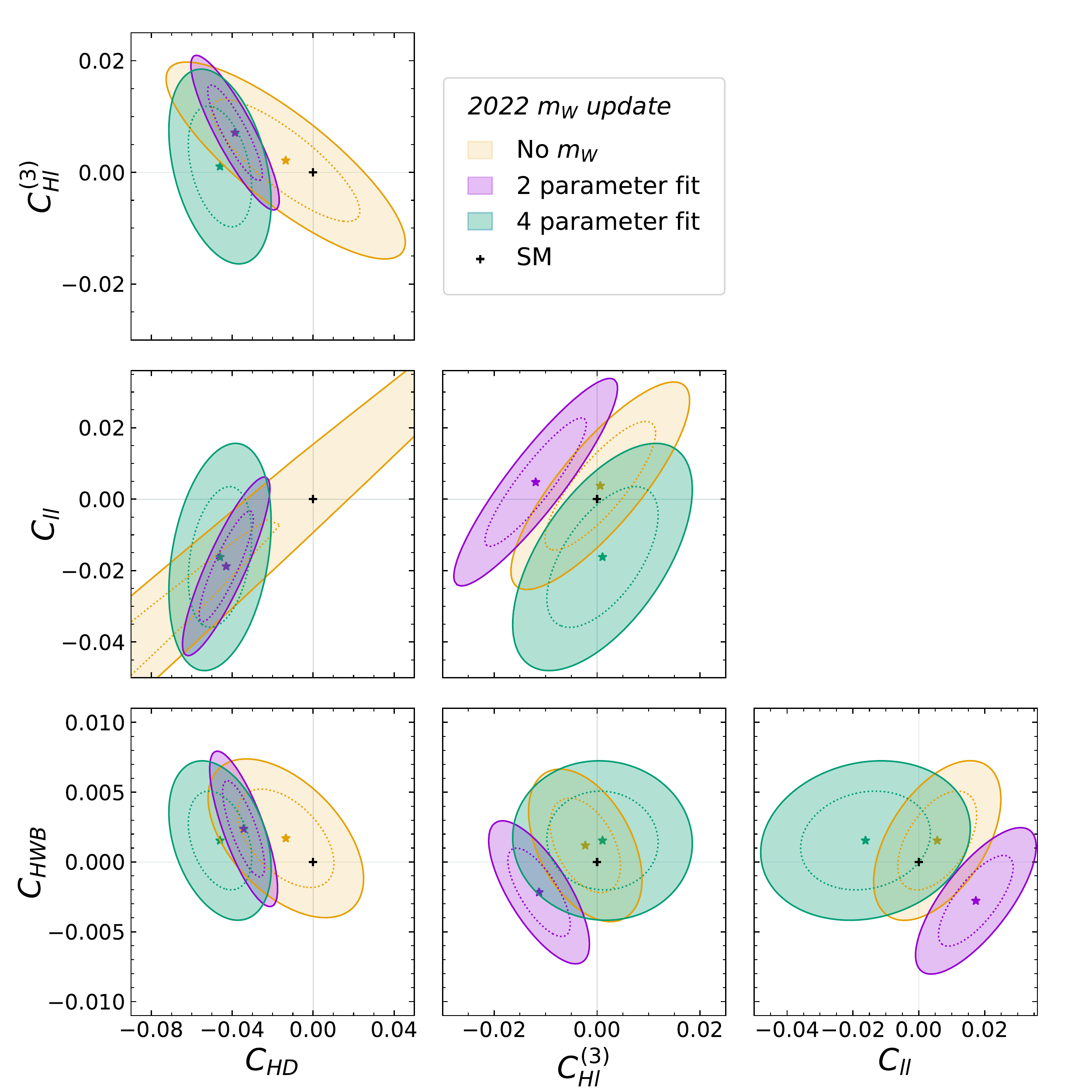}
\caption{
\label{fig:ellipses}
\it Planes of pairs of the coefficients of operators that can affect $m_W$,
comparing the constraints obtained when omitting $m_W$ measurements from the fit (beige),
marginalising over just the four operators that affect $m_W$ (green),
and marginalising over only the two operators in each plane (purple).}
\end{figure}

We now turn to a comparison between indirect and direct measurements of the $W$ mass to assess their compatibility when allowing for different combinations of operator coefficients. Fig.~\ref{fig:SMEFTmW} displays the best fit value and 1-$\sigma$ range of $m_W$ in a variety of SMEFT fits to all combinations of operators entering linearly in $m_W$, including from 1 to 4 operator coefficients, as well as a fit to all 20 operator coefficients. The grey points represent the results omitting $m_W$ from the SMEFT fit, while the red points are for the SMEFT fit including $m_W$. The green band is the SM prediction the input parameters shown in (\ref{inputs}), and the yellow band is the current world average of $m_W$ experimental measurements. Overlap of a grey uncertainty with a red one indicates compatibility at the 1-$\sigma$ level of the latest updated fit with prior data excluding $m_W$. We see that, especially before including an $m_W$ measurement, $C_{HD}$ is the least constrained of the single-parameter scenarios, which enables any fit to subsets of operators involving $C_{HD}$ to find a best fit value of $m_W$ compatible with {our} world average determination. All other subsets excluding $C_{HD}$ are more strongly constrained by data other than $m_W$ and so have a best fit $m_W$ that is pulled down accordingly. Finally, we note that subsets involving $C_{HD}$ and $C_{ll}$ together are pulled to much larger values with large uncertainties if the direct measurement of $m_W$ is not
included in the fit, indicating an almost flat direction that is lifted by the inclusion of $m_W$. Low-energy measurements sensitive to $C_{ll}$ may also break this degeneracy. These have been studied, e.g., in Refs.~\cite{Falkowski:2015krw, Falkowski:2017pss}, but are currently not fully implemented in {\tt Fitmaker}, though we study the impact of low-energy measurements related to unitarity of the CKM matrix in Appendix~\ref{app:CKM}.

We display in Fig.~\ref{fig:ellipses} the 68\% and 95\% CL constraints on pairs of the coefficients of operators capable of modifying $m_W$
that are obtained in a fit using all measurements of $m_W$ including that from CDF, by marginalising over the coefficients of the four operators that affect $m_W$ (green), including just the two operators in each plane (purple), and comparing the results when omitting all $m_W$ measurements (beige). We see in every plane that the 2-operator fit is more constraining than the 4-operator fit.
The constraints also strengthen significantly when including $m_W$. In particular, a flat direction between $C_{ll}$ and $C_{HD}$ is lifted and we see that $C_{HD}$ is bounded away from zero
in each of the three planes where it features, whereas the other operator coefficients may be consistent with zero. However, the SM lies outside the parameter
ellipses in all the planes involving $C_{HD}$.

\FloatBarrier

\section{Probing Single-Field Extensions of the Standard Model}
\label{sec:SFESM}

We now analyse whether our fit favours any particular UV completions of the SM by analysing single-field extensions of the SM Lagrangian, updating the analysis presented in~\cite{Ellis:2020unq}.
The single-field extensions of the SM that contribute to dimension-6 SMEFT operator coefficients at tree level have been catalogued in~\cite{deBlas:2017xtg}, and
we list in Table~\ref{table:fields} the models that can contribute to $m_W$.
The expressions for the dimension-6 operator coefficients generated by these single-field models at the tree-level
are given in Table~\ref{tab:single-field_models}, which is taken directly from~\cite{deBlas:2017xtg}, assuming that only a single coupling to the Higgs is present.
We focus here on tree-level extensions of the SM, since models modifying $m_W$ at loop level are likely to involve particles below the TeV scale, for which the SMEFT approach is questionable.

%We now analyse whether our fit favours any particular single-field extensions of the SM.
%These extensions have been catalogued in~\cite{deBlas:2017xtg}, see Table~\ref{table:fields}
%for the single-field models that can contribute to $m_W$.
%The expressions for the dimension-6 operator coefficients generated by single-field models at the tree level that are relevant for our analysis
%are given in Table~\ref{tab:single-field_models}, assuming only a single coupling to the Higgs is present.
There are no single-field models that contribute at tree level to $C_{HWB}$, and only $S_1$
contributes to $C_{\ell \ell}$. Three single-field models contribute to ${C}_{HD}$,
and four to ${C}^{(3)}_{H \ell}$. In both of the latter cases, these single-field models
also contribute to other operator coefficients, and their contributions vary between models.
As we discuss below, these effects open up the possibility in principle of
discriminating between the
different CDF-friendly single-field extensions of the SM. We note that model $S_1$
generates a negative contribution to $C_{\ell \ell}$ (due to the antisymmetry of the $S_1$ Yukawa matrix in flavour space~\cite{deBlas:2014mba, Felkl:2021qdn}) whereas the data prefer a positive value,
that $\Sigma$ and $\Sigma_1$ generate
positive contributions to $C^{(3)}_{H \ell}$ whereas the data prefer a negative value, and that models $W$ and $B_1$ generate positive contributions to $C_{HD}$ whereas the data again prefer a negative value. For this reason, these models
do not improve on the SM fit. On the other hand,
models $N$ and $E$ generate the preferred sign of $C^{(3)}_{H \ell}$ and models $\Xi$, $B$
and $W_1$
generate the preferred sign of $C_{HD}$, so these models can improve on the SM fit.

\begin{table}[h]
{%\tiny
\begin{center}
\begin{tabular}{| c || c ||c|c|c||c|}
\hline %\hline
    Model &    Spin &  SU(3) &   SU(2) &     U(1) &    Parameters  \\
\hline \hline
 %          $S$ &      &    &       &           &     &    $-\frac{1}{2}$ &       &       &        \\
  %        \hline
        $S_{1}$ &   0   &  1  &   1    &    1    &  ($M_S$, $ \kappa_{S}$)         \\
       \hline
       $\Sigma$ &   $\frac{1}{2}$   &  1  &   3 &  0 &    ($M_\Sigma$,$\lambda_{\Sigma}$)       \\
       \hline
      $\Sigma_{1}$ &  $\frac{1}{2}$   &  1  &   3 &  -1 &    ($M_{\Sigma_1}$,$\lambda_{\Sigma_1}$)       \\
      \hline
           $N$ &  $\frac{1}{2}$   &  1  & 1 &  0 &   $(M_N, \lambda_N)$          \\
           \hline
           $E$ &  $\frac{1}{2}$  &  1  & 1 &  -1 &   $(M_E, \lambda_E)$            \\
           \hline
     $B$ &    1 & 1  &  1    &  0      &   $(M_B,\hat{g}_{H}^{B})$  \\
           \hline
     $B_{1}$ &    1 & 1  &  1    &  1      &   $(M_{B_1}, \lambda_{B_1})$  \\
         \hline
 $\Xi$ &   0 & 1   &  3     &   0     &  ($M_\Xi$,$\kappa_{\Xi}$)    \\
                   \hline
         $W_{1}$ &  1 & 1  &   3    &  1  &    ($M_{W_1}$,$\hat{g}_{W_{1}}^{\phi}$)   \\
         \hline
         $W$ &  1 & 1  &   3    &  0  &    ($M_{W}$,$\hat{g}_{W}^{H}$)   \\
         \hline
%          $\varphi$ &     &   &      &       &      &      &    $-y_{\tau}$ &    $-y_{t}$ &    $-y_{b}$  \\
 %  \hline
  %     \small{$\{B,B_{1} \}$} &     &   &         &    &    &     $-\frac{3}{2}$ &      $-$ $y_{\tau}$ &      $-$ $y_{t}$ &     $-$ $y_{b}$  \\
   %\hline
    %   \small{$\{Q_{1}, Q_{7} \}$} &     &   &      &    &    &      &      &      $y_{t}$ &      \\
%\hline
\end{tabular}
\end{center}
} %end \tiny
\caption{\it The single-field extensions
listed in the first column can make tree-level contributions to $m_W$.
They have the quantum numbers listed in the following three columns,
and the notations for their masses and couplings are given in the last column.
}
\label{table:fields}
\end{table}

\setlength{\tabcolsep}{4.5pt} % Default value: 6pt
\begin{table}[h]
{%\tiny
\begin{center}
\begin{tabular}{| c || c |c|c|c|c|c|c|c|c|c|c|c|c|}
\hline %\hline
    Model &    $C_{HD}$ &  $C_{ll}$ &    $C_{Hl}^{(3)}$ &     $C_{Hl}^{(1)}$ &     $C_{He}$ &   $C_{H \Box}$ &     $C_{\tau H}$ &     $C_{tH}$ &     $C_{bH}$  \\
\hline \hline
 %          $S$ &      &    &       &           &     &    $-\frac{1}{2}$ &       &       &        \\
  %        \hline
\lgr        $S_{1}$ &      &  -1 &       &        &       &             &       &       &         \\
       \hline
 \lgr      $\Sigma$ &      &    &   $\frac{1}{16}$ &  $\frac{3}{16}$ &          &       &    $\frac{y_{\tau}}{4}$ &       &       \\
       \hline
 \lgr     $\Sigma_{1}$ &     &   &   $\frac{1}{16}$ &  $-\frac{3}{16}$ &          &      &  $\frac{y_{\tau}}{8}$ &      &      \\
      \hline
           $N$ &      &    &   $-\frac{1}{4}$ &     $\frac{1}{4}$ &       &              &       &       &         \\
           \hline
           $E$ &      &    &   $-\frac{1}{4}$ &    $-\frac{1}{4}$ &          &       &      $\frac{y_{\tau}}{2}$ &       &        \\
           \hline
%     $\Delta_{1}$ &      &      &    &     &    $\frac{1}{2}$ &       &      $\frac{y_{\tau}}{2}$ &       &        \\
 %    \hline
  %    $\Delta_{3}$ &      &        &    &     &  $-\frac{1}{2}$ &       &      $\frac{y_{\tau}}{2}$ &       &       \\
   %   \hline
 \lgr        $B_{1}$ &    $1$ &   &      &        &    &    $-\frac{1}{2}$ &    $-\frac{y_{\tau}}{2}$ &    $-\frac{y_{t}}{2}$ &    $-\frac{y_{b}}{2} $  \\
         \hline
          $B$ &    $-2$ &   &      &        &    &     &    $-y_{\tau}$ &    $-y_{t}$ &    $-y_{b} $  \\
         \hline
 $\Xi$ &   $-2 \left(\frac{1}{M_\Xi}\right)^2$ &    &       &        &     &     $\frac{1}{2} \left(\frac{1}{M_\Xi}\right)^2$ &     $y_{\tau}\left(\frac{1}{M_\Xi}\right)^2$ &    $y_{t}\left(\frac{1}{M_\Xi}\right)^2$ &     $y_{b}\left(\frac{1}{M_\Xi}\right)^2$  \\
                   \hline
         $W_{1}$ &  $-\frac{1}{4}$&   &       &    &    &  $-\frac{1}{8}$ &  $-\frac{y_{\tau}}{8}$ &  $-\frac{y_{t}}{8}$ &  $-\frac{y_{b}}{8} $  \\
         \hline
\lgr         $W$ &  $\frac{1}{2}$&   &       &    &    &  $-\frac{1}{2}$ &  $-y_{\tau}$ &  $-y_{t}$ &  $-y_{b} $  \\
         \hline
%          $\varphi$ &     &   &      &       &      &      &    $-y_{\tau}$ &    $-y_{t}$ &    $-y_{b}$  \\
 %  \hline
  %     \small{$\{B,B_{1} \}$} &     &   &         &    &    &     $-\frac{3}{2}$ &      $-$ $y_{\tau}$ &      $-$ $y_{t}$ &     $-$ $y_{b}$  \\
   %\hline
    %   \small{$\{Q_{1}, Q_{7} \}$} &     &   &      &    &    &      &      &      $y_{t}$ &      \\
%\hline
\end{tabular}
\end{center}
} %end \tiny
\caption{\it Operators generated at the tree level by the single-field extensions
listed in the first column, which can make tree-level contributions to $m_W$.
%Each extension depends on a single coupling (see Table~\ref{table:fields}) as well as a new physics mass-scale $M$.
The coefficients of the operators are
given by the squares of the corresponding coupling
divided by the corresponding $M^2$, with the exception of an extra factor of $1/m_\Xi^2$
in the case of the $\Xi$ field, as noted in the Table. We  denote the top, bottom and $\tau$ Yukawa couplings by
$y_{t}$, $y_{b}$ and $y_{\tau}$, respectively. Models that contribute to $m_W$ in such a way that they cannot improve upon the SM fit are greyed out.
%, $v$ denotes the electroweak scale and $\alpha_{s}$ is the strong coupling.
}
\label{tab:single-field_models}
\end{table}
\setlength{\tabcolsep}{6pt} % Default value: 6pt

\begin{figure}[t!]
\centering
\includegraphics[width=0.85\textwidth]{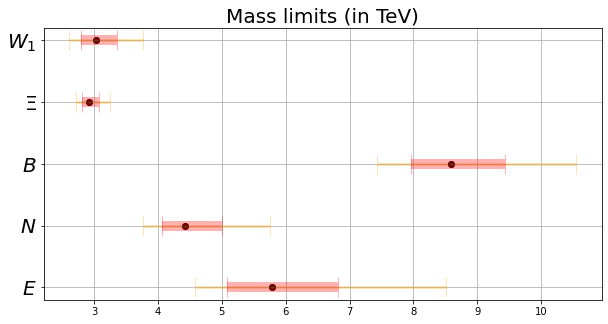}
\caption{\it The horizontal bars show the mass limits (in TeV) at the 68 and 95\% CL for the models described in Table~\ref{table:fields},
setting the corresponding couplings to unity. A larger (smaller) coupling would increase (decrease) the mass ranges proportionally. The coupling limits obtained when setting the mass to 1 TeV are listed in Table~\ref{tab:modelparams}.}
\label{fig:1Dlimits}
\end{figure}

Fig.~\ref{fig:1Dlimits} displays the constraints we find on the single-field extensions of the SM catalogued in Table~\ref{table:fields} that can increase $m_W$ above its SM value.
The salmon and ochre bars show the preferred mass ranges (in TeV) for these models
at the 68\% and 95\% CL, respectively, setting the corresponding model couplings to
unity. The mass ranges would scale linearly with the magnitudes of the couplings. Numerical results are collected in Table~\ref{tab:modelparams},
where we also quote the 68\% CL ranges of the couplings assuming that the masses
of the additional field are 1~TeV.
The rows of the Table are ordered according to decreasing values of the pulls. In the case of the $\Xi$ field, the relevant coupling, $\kappa_{\Xi}$, has dimensions of mass, which explains the additional factor of $M_{\Xi}^{-2}$ in the corresponding entries of Table~\ref{tab:single-field_models}. Here we have made the simplifying choice of fixing $\kappa_\Xi=$1 TeV, although we note that another, equally simple choice could be to set $\kappa_\Xi=M_\Xi$, in which case the results become identical to those of the $B$ model.

The three single-field extensions of the SM that fit best the CDF and other data are models $W_1$, $B$ and $\Xi$, which are
all SU(3) singlets and SU(2) triplets, but differ in their spins and hypercharge (spin 1 with unit hypercharge, spin 1 with zero hypercharge, and spin 0 with zero hypercharge,
respectively). Each of these models exerts a pull of about 6.4 relative to the SM. The next best model is $N$, which is a singlet fermion, also known as a sterile neutrino or heavy neutral lepton,
which is a zero-hypercharge singlet of both SU(3) and SU(2).
%, and $S_1$, which is a singlet scalar with non-zero hypercharge.
This model exerts a
pull of about 5 relative to the SM. Finally, we note that model $E$, which is a singlet fermion with non-zero hypercharge,
exerts a pull of about 3.5. These are the models in Table~\ref{tab:single-field_models} that generate non-zero
coefficients for either $C^{(3)}_{Hl}$ or $C_{HD}$ with the negative sign that is indicated by the individual fits
in Fig.~\ref{fig:flavour_universal_EWdiBH}. The other single-field extensions do not improve upon the SM fit,
either because they do not contribute to either of these operator coefficients,
or because their contributions have the disfavoured sign.
Table~\ref{tab:modelparams} lists the central values and 68\%  and 95\% CL ranges of the masses of the extra fields
that give better fits than the SM, assuming that their couplings are unity,
and the 68\% CL ranges of their couplings, assuming a mass of 1~TeV. We see that
if the coupling in one of these models is of order unity the mass is large
enough for the leading-order SMEFT analysis employed here to be consistent, and
this would probably be the case even if the coupling were ${\cal O}(0.1)$.

\begin{table}[h]
%\begin{adjustwidth}{-0.5in}{-0.55in}% adjust the L and R margins
{%\small
\begin{center}
\begin{tabular}{| c || c ||c |c|c||c|}
\hline %\hline
Model &  Pull &  Best-fit mass &    1-$\sigma$ mass &  2-$\sigma$ mass &  1-$\sigma$ coupling$^2$  \\
 &   &   (TeV) & range (TeV) & range (TeV) &  range  \\\hline \hline
        $W_1$ & 6.4 & 3.0 & [2.8, 3.6]  & [2.6, 3.8]   & [0.09, 0.13]\\
        $B$ & 6.4  & 8.6  & [8.0, 9.4] & [7.4, 10.6]   & [0.011, 0.016] \\
        $\Xi$ & 6.4  & 2.9  & [2.8, 3.1] & [2.7, 3.2]   & [0.011, 0.016] \\
        $N$ & 5.1  & 4.4  & [4.1, 5.0]      & [3.8, 5.8]   & [0.040, 0.060] \\
%        $S_1$ & 4.8  & 6.6 & [5.9, 7.4]  &  [5.4, 8.5]   & [0.018, 0.028] \\
        $E$ & 3.5  & 5.8  & [5.1, 6.8]      &  [4.6, 8.5]   & [0.022, 0.039] \\
        \hline
\end{tabular}
\end{center}
} %end \tiny
%\end{adjustwidth}
\caption{\it Single-field models that can improve on the SM fit when the CDF measurement of $m_W$ is included,
showing their respective pulls, the best-fit masses and their 1- and 2-$\sigma$ ranges assuming unit couplings,
and the 1-$\sigma$ coupling ranges assuming masses of 1~TeV.}
\label{tab:modelparams}
\end{table}

We have analysed the effect on this analysis of applying the PDG recommendation~\cite{ParticleDataGroup:2016lqr} for rescaling the
stated uncertainties when combining measurements that are only poorly consistent. The rescaling factor for combining the various $m_W$ measurements
shown in Fig.~\ref{fig:mW} is a factor $\sim 2$, and leads to the revised numbers for the preferred parameter
ranges for single-field extensions of the SM shown in Table~\ref{tab:modelparamsPDG}. In general, the pulls for the single-field extensions are
reduced, as one would expect. The best-fit masses for the bosonic models $W_1, B$ and $\Xi$ are not changed substantially, increasing by $\sim$1$\sigma$
in each case, but there are larger changes in the best-fit masses for the fermionic models $N$ and $E$. The changes in the 1- and 2-$\sigma$ parameter
ranges for the different models reflect these effects.

\begin{table}[h]
%\begin{adjustwidth}{-0.5in}{-0.55in}% adjust the L and R margins
\begin{center}
\begin{tabular}{| c || c ||c |c|c||c|}
\hline %\hline
Model &  Pull &  Best-fit mass &    1-$\sigma$ mass &  2-$\sigma$ mass &  1-$\sigma$ coupling$^2$  \\
 &   &   (TeV) & range (TeV) & range (TeV) &  range  \\\hline \hline
        $W_1$ & 2.8 & 3.3 & [2.9, 4.3]  & [2.6, 3.8]   & [0.06, 0.12]\\
        $B$ & 2.8  & 10.0  & [8.2, 12.0] & [7.4, 10.6]   & [0.007, 0.015] \\
        $\Xi$ & 2.8  & 3.1  & [2.9, 3.5] & [2.7, 3.3]   & [0.007, 0.015] \\
        $N$ & 2.1  & 6.5  & [5.1, 8.6]      & [4.6, 20.8]   & [0.013, 0.038] \\
        $E$ & 0.6  & 13.6  & [8.1, 29.7]      &  [6.3, $\infty$)   & [0.001, 0.015] \\
        \hline
\end{tabular}
\end{center}
%\end{adjustwidth}
        \caption{\it As Table~\ref{tab:modelparams}, but applying the PDG recommendation~\cite{ParticleDataGroup:2016lqr} for rescaling the
stated uncertainties when combining measurements that are only poorly consistent.}
\label{tab:modelparamsPDG}
\end{table}

\section{Prospects for Direct Detection of New Particles at the LHC}
\label{sec:detect}

The single-field extension that gives the best fit, $W_1$, is an isospin triplet vector boson with non-zero hypercharge,
which is not a common feature of unified gauge theories. The next-best fit introduces $B$, a singlet vector boson with zero hypercharge, commonly known as a $Z^\prime$, which was shown previously to be able to increase the $W$ mass~\cite{Allanach:2021kzj, Alguero:2022est}, and $\Xi$, a spin-zero isotriplet boson
with zero hypercharge, which is a type of field that appears in some extended Higgs sectors
(see, e.g.,~\cite{Diessner:2019ebm}). Among the other fields that could
improve on the SM fit, $N$ and $E$ are singlet fermions with zero and non-zero hypercharges, respectively, and the $N$ field
could be identified with a singlet heavy neutral lepton.
%Finally, the $S_1$ field is a singlet scalar with non-zero hypercharge.

The particles that could best explain the new $W$ mass measurement, the vector and scalar triplets $W_1$ and $\Xi$ would have masses around 3 TeV for a coupling of ${\cal O}(1)$ and 1 TeV, respectively. In such a case these particles would be kinematically accessible at the LHC, with a guaranteed production mechanism via their electroweak couplings. Moreover,
as the SMEFT constrains only the ratio of coupling to mass, masses below the TeV scale would be favoured in a weak coupling scenario.  On the other hand,  the best fit for the vector $B$ mass is around 8.6 TeV for an ${\cal O}(1)$ coupling, making this option less interesting from the direct search point of view.

We note that, among all the possible interactions these new particles could have, our analysis is sensitive only to the coupling to the electroweak symmetry-breaking sector, e.g., the triplet coupling $\kappa_\Xi \, H^\dagger \, \Xi^a \sigma^a H$. Heavy triplets would be produced in pairs, or produced
singly  in vector-boson fusion (VBF) with a suppression by the ratio of the vevs of the triplet and doublet, $v_t/v$. The triplet would then decay to bosons if kinematically allowed, see, e.g., Refs.~\cite{Chabab:2018ert} and~\cite{Bandyopadhyay:2020otm} for studies of the collider
phenomenologies of real and complex scalar triplets.
In more complete scenarios such as supersymmetry or the Georgi-Machacek model~\cite{Georgi:1985nv}, the triplet would be accompanied by other new particles and exhibit a richer phenomenology, including a candidate for dark matter, see, e.g., Ref.~\cite{Bell:2020hnr} for a recent study in the context of supersymmetry.

There have been various searches for heavy charged Higgses that apply to the $\Xi$ scenario. For example, CMS has performed a VBF search in $WW$ and $WZ$ final states with the full Run~2 dataset~\cite{CMS:2021wlt}. Their search is sensitive to  $\sigma \times  Br\sim$ $\cal O$(few fb) in the mass range 1-3 TeV, right in the ballpark of the preferred region from the SMEFT fit. The phenomenology of $B$, also known as a $Z'$, is well understood, but searches often focus on possible couplings to light fermions through their $U(1)^\prime$ charges and/or mixing with the SM gauge bosons. On the other hand, if we rely just on the coupling to the Higgs sector, the primary phenomenological consequence is the presence of $Z-Z^\prime$ mass-mixing. This leads to the heavy gauge boson inheriting the couplings of the $Z$-boson, suppressed by the mixing angle $\sim\hat{g}_H^B v^2/M_B^2$~\cite{Babu:1997st}. Its production and decay modes would therefore be similar to the $Z$ itself. We estimate that, for $\hat{g}_H^B\sim\mathcal{O}(1)$, current LHC searches for dilepton resonances are only sensitive to $M_B\sim1.5$ TeV~\cite{ATLAS:2019erb}, suggesting that it will be challenging for the LHC to probe the viable parameter space of this simplified scenario.

Interpreting existing direct collider searches for $W_1$, the vector triplet with hypercharge $Y=1$, is not as straightforward as for the $\Xi$ particle. Whereas there have been
studies on the phenomenology of $Y=0$ vector triplets, see e.g. Refs.~\cite{Lizana:2013xla,Pappadopulo:2014qza}, the translation of these limits into the new scenario would require developing a tailored search strategy. Currently, LHC searches are interpreted within various benchmark scenarios~\cite{CMS:2021fyk}: a scenario where the couplings to fermions and gauge bosons are of the same order, a mass-suppressed fermion couplings scenario as in composite Higgs Models, and a purely-bosonic coupling scenario. In the first two cases, searches for a $ZH$ final state are nowadays sensitive to $\sigma \times  Br\lesssim$ $\cal O$(1 fb) in the mass range 2-5 TeV. On the other hand, the case of a purely-bosonic state is similar to the  situation described above for the scalar triplet, where the production is via VBF and the final state contains massive bosons.

\section{Conclusions}
\label{sec:conx}

The recent CDF measurement of $m_W$ poses a strong challenge, not only to the Standard Model, but also to many well-studied
extensions such as supersymmetry~\cite{Bagnaschi:2022qhb}. However, we have shown that it is compatible with a general leading-order
dimension-6 SMEFT analysis: new physics parametrised by dimension-6 operators can generically account for a large enough shift in $m_W$ without any significant tension with other electroweak precision, Higgs and diboson data.

Table~\ref{table:scenarios} summarises the qualities of SMEFT fits that we find under various assumptions,
either without using any $m_W$ measurement, or using only the measurements prior to
the recent CDF measurement, or combining the CDF measurement with previous measurements.
%, either naively or (in parentheses) following the PDG prescription for expanding the combined error when different measurements are in tension (a factor $\sim 2.1$ for the $m_W$ measurements).
%In each scenario, we present also results including the Cabibbo-Kobayashi-Maskawa unitarity constraint that $\Delta_{CKM} \equiv |V_{ud}|^2 + |V_{us}|^2 - 1 = - 0.0015 \pm 0.0007 = 2 \frac{v^2}{\Lambda^2} \left[C_{Hq}^{(3)} - C_{H \ell}^{(3)} + C_{\ell \ell} - C_{\ell q}^{(3)} \right]$ when dimension-6 SMEFT operators are included~\cite{Cirigliano:2022qdm}.
We present the qualities of fits using the full set of 20 dimension-6 SMEFT operators
and restricting to only the 4 operators that can increase $m_W$ relative to its SM value. We see that in all cases the $\chi^2/{\rm dof} < 1$
and the $p$-values are $> 0.5$, indicating high levels of consistency with the data.

\begin{table}[t]
{%\tiny
\begin{center}
\begin{tabular}{| c | c |c|c|c|c|c|}
\hline %\hline
    EWPO, H &    Previous  & Combined& Parameter & N$_{\rm dof}$ & $\chi^2/{\rm dof}$ & $p$-value \\
    diboson      &  $m_W$   & $m_W$           &  Count   &   &       &             \\
\hline \hline
    $\checkmark$ &              & &                20   & 182  &    0.92    &    0.76         \\
%    $\checkmark$ &              & & $\checkmark$ &  20   & 183  &    0.94    &    0.71         \\
\hline
    $\checkmark$ & $\checkmark$ & &                20   & 185  &    0.93    &    0.75         \\
%    $\checkmark$ & $\checkmark$ & & $\checkmark$ &  20   & 186  &    0.93    &    0.74      \\
\hline
%$\checkmark$      &  & $\checkmark$  &      &          &  20   & 185  &    0.97    &    0.59         \\
 %   $\checkmark$      &  & $\checkmark$  &      &  $\checkmark$  &  20   & 186  &    0.98    &    0.56         \\
%\hline
    $\checkmark$ & & $\checkmark$ &                20   & 185   &   0.97     &    0.59        \\
%    $\checkmark$ & & $\checkmark$ & $\checkmark$ &  20   & 186   &   0.98     &    0.56        \\
\hline \hline
    $\checkmark$ &              & &                4    & 198   &    0.93    &    0.76         \\
%    $\checkmark$ &              & & $\checkmark$ &  4    & 199   &    0.93    &    0.74         \\
\hline
    $\checkmark$ & $\checkmark$ & &                4    & 201   &    0.93    &    0.75      \\
 %   $\checkmark$ & $\checkmark$ & & $\checkmark$ &  4    & 202   &    0.93    &    0.75      \\
\hline
%$\checkmark$      &  & $\checkmark$  &      &          &  4   & 201  &    0.97    &    0.60         \\
%    $\checkmark$      &  & $\checkmark$  &      &  $\checkmark$  &  4   & 221  &    0.97    &    0.62         \\
%\hline
    $\checkmark$ & &        $\checkmark$       &     4   & 201   &    0.97     &    0.60        \\
 %   $\checkmark$ & & $\checkmark$  & $\checkmark$ &   4   & 202   &    0.97     &    0.62        \\
\hline\end{tabular}
\end{center}
} %end \tiny
\caption{\it Fits to the electroweak, Higgs and diboson data either without any $m_W$ measurement, or with the measurements prior to
the recent CDF measurement, or with the combination of the CDF and previous measurements, using either the
full set of 20 SMEFT operators or only the 4 operators that can increase $m_W$ relative to its SM value.
}
\label{table:scenarios}
\end{table}

We have also used the SMEFT to show that the CDF and other $m_W$ measurements can be accommodated within several single-field extensions of the Standard Model
with new particles whose masses are in the TeV range for couplings of order unity, in which case the SMEFT approach is self-consistent. We find the strongest pulls for an electroweak triplet, either scalar or vector with zero or unit hypercharge respectively, a singlet $Z^\prime$ vector boson, followed by a singlet heavy neutral lepton. The LHC searches made so far have not
excluded particles with masses and couplings in the favoured ranges, but there are prospects for LHC Run~3 and HL-LHC
that merit more detailed study.

\subsubsection*{Note Added}
{
As we were completing this paper, several
papers~\cite{Fan:2022dck,Zhu:2022tpr,Lu:2022bgw,Athron:2022qpo,Yuan:2022cpw,Strumia:2022qkt,Yang:2022gvz,deBlas:2022hdk}
appeared that also discuss the impact of the new $m_W$ measurement on new physics scenarios.

Ref.~\cite{deBlas:2022hdk} is the most similar in spirit to our work, in that it also reports results from fits to oblique observables and a SMEFT fit,
though only to electroweak precision observables and without specific model interpretations.
Ref.~\cite{Strumia:2022qkt} also discusses oblique observables and considers
interpretations invoking heavy $Z'$ bosons, little-Higgs models or higher-dimensional geometries.
Refs~\cite{Fan:2022dck,Lu:2022bgw,Zhu:2022tpr} discuss two-Higgs doublet models, Ref.~\cite{Athron:2022qpo} proposes a leptoquark interpretation that also accommodates
$g_\mu - 2$, Ref~\cite{Yuan:2022cpw} considers various models with light degrees of freedom, and Ref.~\cite{Yang:2022gvz} discusses a supersymmetric interpretation.
}

%\newpage

\acknowledgments
We thank Tobias Felkl for valuable correspondence on the sign of the $S_1$ contribution. The work of J.E. was supported in part by the UK STFC via grants ST/P000258/1
and ST/T000759/1, and in part by the Estonian Research Council via grant MOBTT5. M.M. is supported by the European Research Council under the European
Union’s Horizon 2020 research and innovation Programme (grant agreement n.950246) and in part by STFC consolidated grant ST/T000694/1. K.M.
is also supported by the UK STFC via grant ST/T000759/1.
V.S. is supported by the PROMETEO/2021/083 from Generalitat
Valenciana, and by PID2020-113644GB-I00 from the Spanish Ministerio de Ciencia e Innovaci\'on.
T.Y. is supported by a Branco Weiss Society in Science Fellowship and partially by
the UK STFC via the grant ST/P000681/1.

%\newpage

\appendix

\section{Constraints from CKM unitarity}
\label{app:CKM}

\begin{figure}[h!]
\centering
\includegraphics[width=0.8\textwidth]{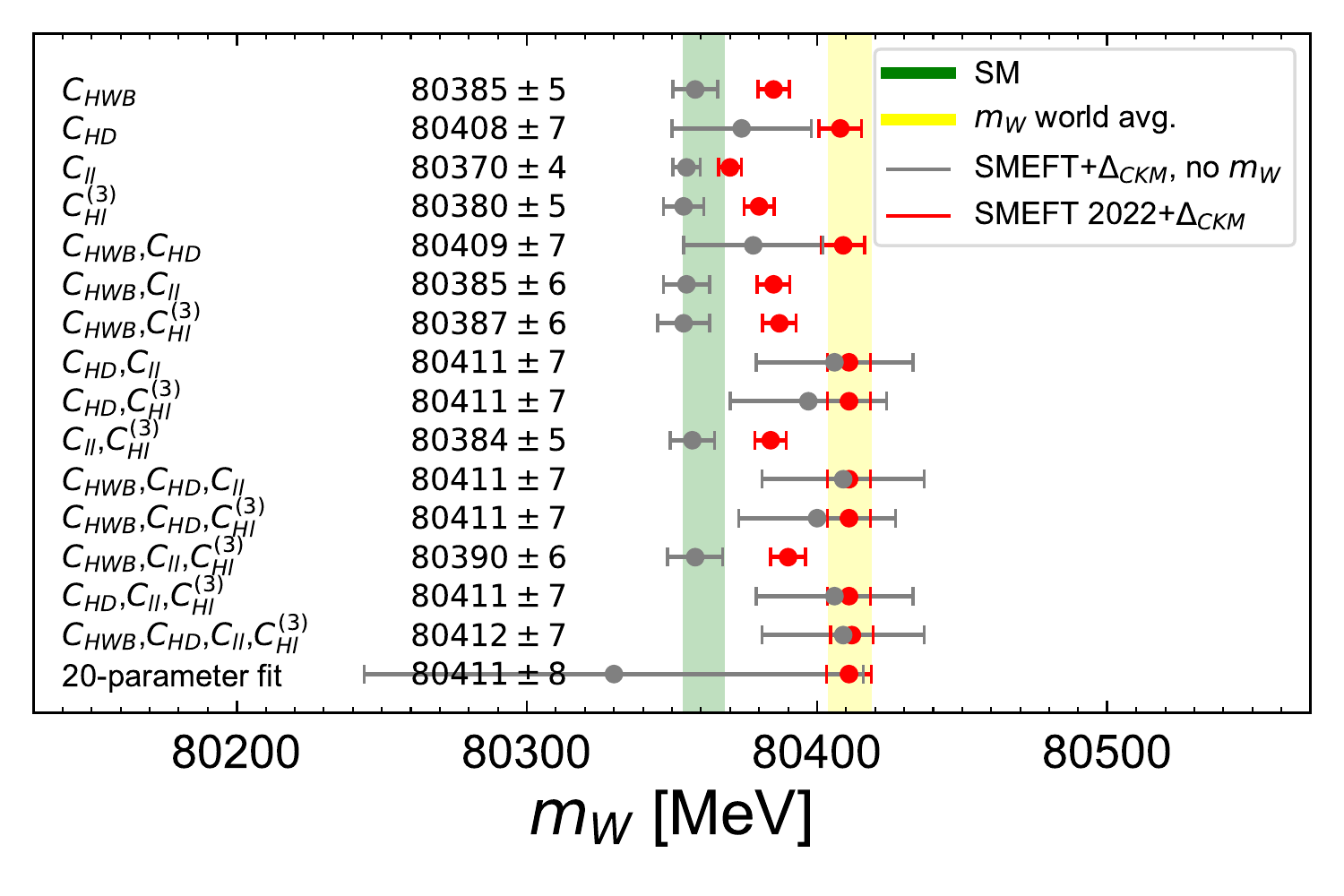}
\caption{
\label{fig:SMEFTsummaryCKM}
\it Values of $m_W$ in fits including all combinations of operators entering linearly in $m_W$, including from 1 to 4 operator ceofficients, as well as a fit to 20 operator coefficients. Results omitting direct measurements of $m_W$ but including $\Delta_{CKM}$ are shown in grey, and results with $\Delta_{CKM}$ and using the current world average of $m_W$ measurements including that by CDF are shown in red. The vertical green band is the SM prediction for $m_W$ based on a SM electroweak fit excuding $m_W$ and $\Delta_{CKM}$, and the yellow band is the current world average of $m_W$ measurements.}
\end{figure}

It was pointed out in Refs.~\cite{Blennow:2022yfm, Cirigliano:2022qdm} that the consistency of $\beta$-decay
measurements with the unitarity of the Cabibbo-Kobayashi-Maskawa (CKM) quark mixing matrix
imposes a significant constraint on one combination of the dimension-6 SMEFT coefficients.
Specifically, the quantity $\Delta_{CKM} \equiv |V_{ud}|^2 + |V_{us}|^2 - 1$ can be expressed as
\begin{equation}
    \Delta_{CKM} = 2 \frac{v^2}{\Lambda^2} \left[C_{Hq}^{(3)} - C_{H \ell}^{(3)} + C_{\ell \ell} - C_{\ell q}^{(3)} \right] \, ,
    \label{CKM}
\end{equation}
and measurements of $0^+ \to 0^+$ nuclear transitions  and kaon decays indicate that
\begin{equation}
    \Delta_{CKM} = -0.0015 \pm 0.0007 \, ,
    \label{CKMconstraint}
\end{equation}
which is consistent with CKM unitarity at the per mille level. In this appendix we discuss the impact of including this measurement in our fit, leaving a more complete study of low-energy measurements to future work.

\begin{figure}[t!]
\centering
\includegraphics[width=0.8\textwidth]{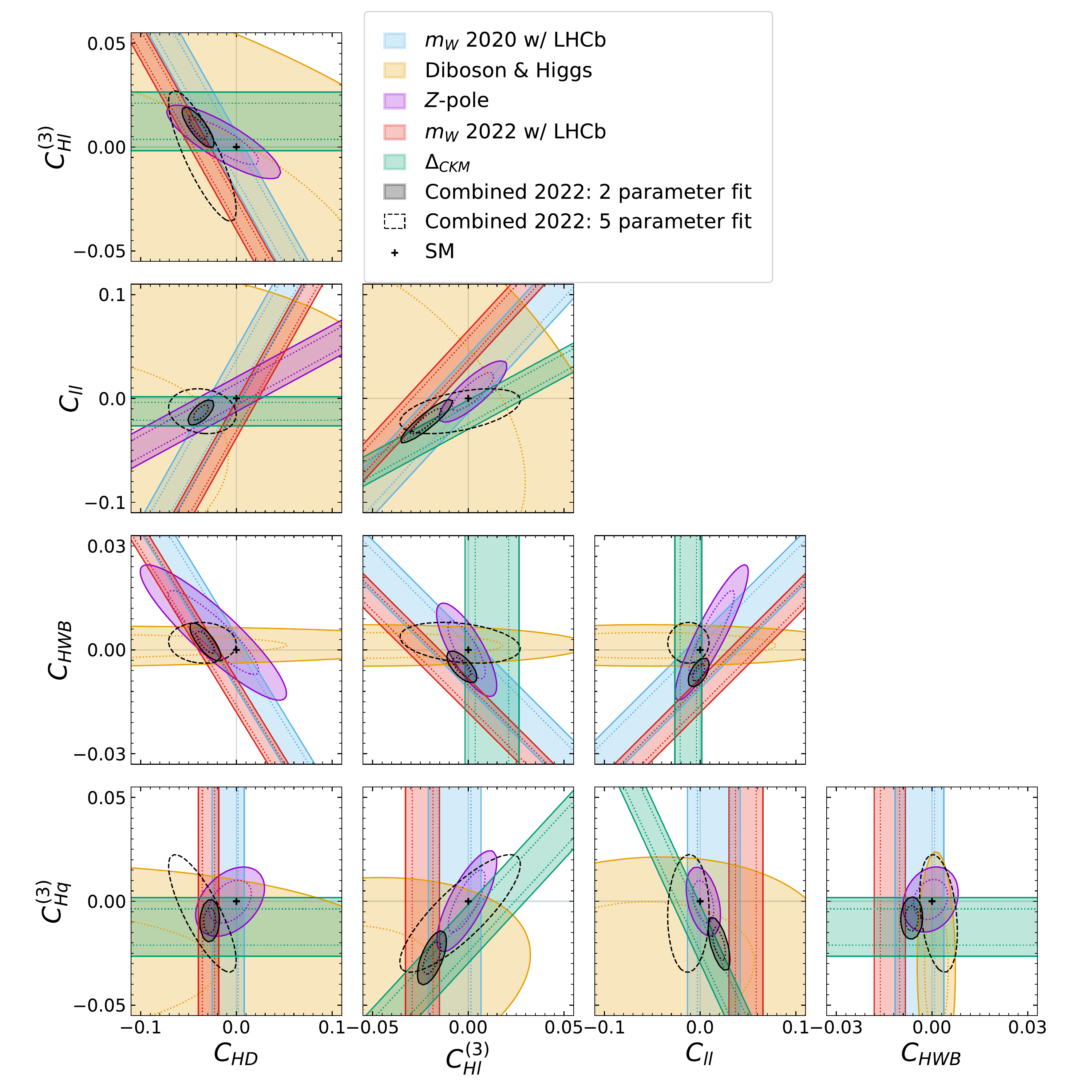}
\caption{
\label{fig:ellipsesCKM}
\it Planes of pairs of coefficients of operators that can affect $m_W$ and $\Delta_{CKM}$, showing the constraints from direct $m_W$ measurements before (blue) and after (red) the recent CDF update, $Z$-pole data (purple), diboson and Higgs data (beige), and $\Delta_{CKM}$ (green). The combined constraints are given by solid and dashed dotted lines for a 2- and 5-parameter fit respectively.}
\end{figure}

We note that two of the operators listed in (\ref{CKM}) that contribute to
$\Delta_{CKM}$ also contribute to $m_W$, namely $C_{H \ell}^{(3)}$
and $C_{\ell \ell}$, whereas the other two operators that contribute to
$\Delta_{CKM}$, namely $C_{Hq}^{(3)}$ and $C_{\ell q}^{(3)}$, do not contribute
to $m_W$. Conversely, whereas $C_{HWB}$ and $C_{HD}$ contribute to $m_W$,
they do not contribute to $\Delta_{CKM}$. The upshot of this summary is that
the CKM unitarity constraint (\ref{CKMconstraint}) is irrelevant for the
coefficients of the operators
${\cal O}_{HWB}$ and ${\cal O}_{HD}$, but is potentially important for
${\cal O}_{H \ell}^{(3)}$ and ${\cal O}_{\ell \ell}$. Two scenarios can be
distinguished, either ${C}_{H \ell}^{(3)} = {C}_{\ell \ell}$, in which
case the constraints on the non-$m_W$ operator coefficients
$C_{Hq}^{(3)}$ and $C_{\ell q}^{(3)}$ can be
considered separately, or there is no cancellation between
${C}_{H \ell}^{(3)}$ and ${C}_{\ell \ell}$, in which case all four of the
operator coefficients in (\ref{CKM}) must be considered together.

\begin{table}[t]
{%\tiny
\begin{center}
\begin{tabular}{| c | c |c|c||c|c|c|c|}
\hline %\hline
    EWPO, H &    Previous  & Combined& $\Delta_{CKM}$ & Parameter & N$_{\rm dof}$ & $\chi^2/{\rm dof}$ & $p$-value \\
    diboson      &  $m_W$   & $m_W$      &           &  Count   &   &       &             \\
\hline \hline
%    $\checkmark$ &              & &              &  20   & 182  &    0.92    &    0.76         \\
    $\checkmark$ &              & & $\checkmark$ &  20   & 183  &    0.94    &    0.71         \\
\hline
%    $\checkmark$ & $\checkmark$ & &              &  20   & 185  &    0.93    &    0.75         \\
    $\checkmark$ & $\checkmark$ & & $\checkmark$ &  20   & 186  &    0.93    &    0.74      \\
\hline
%$\checkmark$      &  & $\checkmark$  &      &          &  20   & 185  &    0.97    &    0.59         \\
 %   $\checkmark$      &  & $\checkmark$  &      &  $\checkmark$  &  20   & 186  &    0.98    &    0.56         \\
%\hline
 %   $\checkmark$ & & $\checkmark$ &              &  20   & 185   &   0.97     &    0.59        \\
    $\checkmark$ & & $\checkmark$ & $\checkmark$ &  20   & 186   &   0.98     &    0.56        \\
\hline \hline
%    $\checkmark$ &              & &              &  4    & 198   &    0.93    &    0.76         \\
    $\checkmark$ &              & & $\checkmark$ &  4    & 199   &    0.93    &    0.74         \\
\hline
%    $\checkmark$ & $\checkmark$ & &              &  4    & 201   &    0.93    &    0.75      \\
    $\checkmark$ & $\checkmark$ & & $\checkmark$ &  4    & 202   &    0.93    &    0.75      \\
\hline
%$\checkmark$      &  & $\checkmark$  &      &          &  4   & 201  &    0.97    &    0.60         \\
%    $\checkmark$      &  & $\checkmark$  &      &  $\checkmark$  &  4   & 221  &    0.97    &    0.62         \\
%\hline
 %   $\checkmark$ &  & $\checkmark$ & &  4   & 201   &    0.97     &    0.60        \\
    $\checkmark$ & & $\checkmark$  & $\checkmark$ &   4   & 202   &    0.97     &    0.62        \\
\hline\end{tabular}
\end{center}
} %end \tiny
\caption{\it
As Table~\ref{table:scenarios}, but including also the $\Delta_{CKM}$ constraint.
%Fits to the electroweak, Higgs and diboson data either without any $m_W$ measurement, or with the measurements prior to
%the recent CDF measurement, or with the combination of the CDF and previous measurements.
%In each case we show results with and without the $\Delta_{CKM}$ constraint, and using either the
%full set of 20 SMEFT operators or only the 4 operators that can increase $m_W$ relative to its SM value.
}
\label{table:scenariosCKM}
\end{table}

Fig.~\ref{fig:SMEFTsummaryCKM} shows the $m_W$ values in grey corresponding to the best fit and 1-$\sigma$ range allowed by our fit to various coefficient subsets of the SMEFT without including direct $m_W$ measurements, but with the constraint from $\Delta_{CKM}$. Comparing to Fig.~\ref{fig:SMEFTmW}, we see that flat directions involving $C_{HD}$ and $C_{ll}$ are lifted by including $\Delta_{CKM}$, giving indirect predictions for $m_W$ compatible with the world average of the direct measurements, which  is represented by the yellow band. This is further illustrated in Fig.~\ref{fig:ellipsesCKM}, where two-dimensional contours corresponding to the constraints from various measurements are plotted for all combinations of the 5 parameters $C_{HD}, C_{Hl}^{(3)}, C_{ll}, C_{HWB}, C_{Hq}^{(3)}$. The flat direction for $C_{ll}$ vs $C_{HD}$ corresponding to the $Z$-pole data in purple can be removed by either including direct $m_W$ measurements, shown before (blue) and after (red) the latest CDF measurement, or by including $\Delta_{CKM}$ in green. We note also in beige the importance of diboson and Higgs data for constraining the $C_{HWB}$ operator. The solid black and dashed black ellipses correspond to the combined bounds for the 2-parameter and marginalised 5-parameter fits respectively.

Turning now to the impact of $\Delta_{CKM}$ on the single-field extensions of the SM studied here, we note that none
of them contribute to $C_{HWB}$, whereas models $B, \Xi$ and $W_1$
contribute to $C_{HD}$ with the appropriate sign to increase $m_W$ and are
not constrained by CKM unitarity. Models $N$ and $E$ also contribute to
$m_W$ with the appropriate sign, and also to
${C}_{H \ell}^{(3)}$ but not ${C}_{\ell \ell}$. Hence the CKM unitarity
constraint is relevant to these models, but it may be satisfied by contributions
from $C_{Hq}^{(3)}$ and $C_{\ell q}^{(3)}$. It seems likely that the
current constraints
on these coefficients from a global SMEFT fit are weak enough to cancel
the ${C}_{H \ell}^{(3)}$ contribution to $\Delta_{CKM}$ to the level needed.

Finally, in Table~\ref{table:scenariosCKM} we summarise the $\chi^2$ per degree of freedom and $p$-values for our 20- and 4-parameter global fits including $\Delta_{CKM}$, both without and together with $m_W$ measurements before and after the latest CDF result.
We again see that in all cases the $p$-values indicate high levels of consistency with the data.

\bibliographystyle{JHEP}
\bibliography{biblio}

\end{document}

%% file: limit_table.tex
\begin{tabular}{|c||c|c|c||c|c|c|}
\hline
 & \multicolumn{3}{c||}{ Individual} & \multicolumn{3}{c|}{Marginalised} \\
 \hline
 SMEFT &  Best fit &   95\% CL &  Scale &  Best fit & 95\% CL &  Scale \\
 Coeff.   & [$\Lambda = 1$~TeV] & range & $\frac{\Lambda}{\sqrt{C}}$ [TeV] & [$\Lambda = 1$~TeV] & range & $\frac{\Lambda}{\sqrt{C}}$ [TeV] \\
\hline \hline
     $C_{HWB}$&-0.01 &[ -0.009, -0.0034 ] & 19.0 & 0.25 &  [ -0.3, +0.81 ]   & 1.3  \tabularnewline\hline
      $C_{HD}$&-0.03 & [ -0.035, -0.019 ] & 11.0 & -0.6 &  [ -1.8, +0.63 ]   & 0.9  \tabularnewline\hline
      $C_{ll}$& 0.02 & [ +0.014, +0.034 ] & 10.0 &-0.05 &[ -0.099, +0.0043 ] & 4.4  \tabularnewline\hline
$C_{Hl}^{(3)}$&-0.01 &[ -0.019, -0.0083 ] & 14.0 &-0.01 & [ -0.11, +0.076 ]  & 3.3  \tabularnewline\hline
$C_{Hl}^{(1)}$&0.00  &[ -0.0045, +0.013 ] & 11.0 & 0.16 &  [ -0.15, +0.47 ]  & 1.8  \tabularnewline\hline
      $C_{He}$& 0.00 &[ -0.015, +0.0071 ] & 9.6  & 0.28 &  [ -0.34, +0.9 ]   & 1.3  \tabularnewline\hline
$C_{Hq}^{(3)}$& 0.00 & [ -0.013, +0.011 ] & 9.1  &-0.05 & [ -0.11, +0.012 ]  & 4.1  \tabularnewline\hline
$C_{Hq}^{(1)}$& 0.01 & [ -0.027, +0.043 ] & 5.4  &-0.07 &  [ -0.2, +0.06 ]   & 2.8  \tabularnewline\hline
      $C_{Hd}$&-0.03 & [ -0.13, +0.072 ]  & 3.1  &-0.44 & [ -0.96, +0.079 ]  & 1.4  \tabularnewline\hline
      $C_{Hu}$& 0.00 & [ -0.075, +0.073 ] & 3.7  &-0.18 &  [ -0.62, +0.26 ]  & 1.5  \tabularnewline\hline
    $C_{HBox}$&-0.27 &   [ -1, +0.47 ]    & 1.2  & -1.1 &    [ -3.2, +1 ]    & 0.69 \tabularnewline\hline
      $C_{HG}$& 0.00 &[ -0.0034, +0.0032 ]& 17.0 &-0.01 & [ -0.026, +0.013 ] & 7.2  \tabularnewline\hline
      $C_{HW}$& 0.00 & [ -0.012, +0.006 ] & 11.0 & 0.18 &  [ -0.33, +0.7 ]   & 1.4  \tabularnewline\hline
      $C_{HB}$& 0.00 &[ -0.0034, +0.002 ] & 19.0 & 0.09 & [ -0.074, +0.24 ]  & 2.5  \tabularnewline\hline
       $C_{W}$& 0.18 & [ -0.072, +0.42 ]  & 2.0  & 0.15 &   [ -0.1, +0.4 ]   & 2.0  \tabularnewline\hline
       $C_{G}$&-0.75 &    [ -4, +2.5 ]    & 0.56 & 1.3  &   [ -6.1, +8.7 ]   & 0.37 \tabularnewline\hline
  $C_{\tau H}$& 0.01 & [ -0.015, +0.025 ] & 7.1  &0.00  & [ -0.017, +0.027 ] & 6.7  \tabularnewline\hline
   $C_{\mu H}$& 0.00 &[ -0.0057, +0.005 ] & 14.0 & 0.00 &[ -0.0056, +0.0052 ]& 14.0 \tabularnewline\hline
      $C_{bH}$&0.00  & [ -0.016, +0.024 ] & 7.1  & 0.02 & [ -0.027, +0.058 ] & 4.8  \tabularnewline\hline
      $C_{tH}$&-0.09 &   [ -1, +0.84 ]    & 1.0  & -2.7 &   [ -8.8, +3.3 ]   & 0.41 \tabularnewline\hline
\end{tabular}